\documentclass[aps,prl,twocolumn,showpacs,superscriptaddress,groupedaddress]{revtex4}  
\usepackage{graphicx,tensor, cleveref}  
\usepackage{dcolumn}   
\usepackage{bm}        
\usepackage{amssymb}   
\usepackage{amsmath}
\usepackage{dsfont}
\usepackage{cancel}
\usepackage{xcolor}

\begin{document}



\title{Orientational Distribution of an Active Brownian Particle: an analytical study}
\author{Supurna Sinha}

\address{Raman Research Institute, Bangalore 560080, India.}

\date{\today}

\begin{abstract}
We use the Fokker Planck equation as a starting point for studying the orientational probability distribution 
of an Active Brownian Particle (ABP) in $(d+1)$ dimensions. This Fokker Planck equation admits an exact solution in series form 
which is, however, unwieldly to use because of poor convergence for short and intermediate times. A truncated version of this series is a reasonable
approximation for long times. 
In this paper, we present an analytical closed form expression, 
which gives a good approximate orientational probability distribution, which is derived  
using saddle point methods for short times. However, it works well even for intermediate times. 
Thus, we have simple analytical forms for the ${\it entire}$ range of time scales for the orientational probability distribution
of an ABP.
Our predictions can be tested against future experiments and simulations probing orientational probability distribution of an ABP.

\end{abstract}

\maketitle



\section{I. Introduction}

The study of active matter has been the focus of current research. 
Active particles are self-propelled particles which generate 
dissipative directed motion by consuming energy from the environment.
There are many examples of active matter in  soft matter and biological systems like 
bacterial run and tumble motion, swimming microbes,
schools of fish, swarms of birds, driven granular matter and so on.

There has been a considerable amount of theoretical, simulational and experimental work in this area both at 
the large scale hydrodynamic level\cite{cristina,sriram} and at the level of single particle dynamics\cite{abhisemi,urna}.
While there has been a fair amount of research in studying the position distribution function of a single active Brownian particle,
there are fewer studies addressing the orientational probability distribution of an ABP.

In this paper we focus on the orientational probability distribution of an ABP
using the Fokker Planck equation as a starting point. 
As is well known, the exact solution to this can be expressed in terms of the eigenvalues and 
eigenfunctions of the Laplacian\cite{zheng,fokkerd}.
Our main results are the following: (a) An  approximate short time closed form analytical expression which, in 
fact turns out to be an effective approximation even at intermediate times.
(b) A long time approximate form which works well at long times.
Thus, we have obtained simple analytical forms for the orientational probability distribution 
for the ${\it entire}$ range of time scales.  

The orientational probability distributions predicted in our analysis can be 
probed experimentally via experiments on Janus particles, for instance\cite{zheng}.

The paper is organized as follows. In Section II 
we discuss the Fokker Planck equation for the orientational dynamics of an active Brownian particle (ABP) in 
$d$ dimensions and present an exact series solution for the distribution in three dimensions. In Section III we discuss 
a short time approximate orientational probability distribution in $(d+1)$ dimensions
and discuss the particular case of three dimensions. 
In Section IV we present a long time approximate form for the orientational probability distribution. 
In Section V we compare the short time and the long time approximate orientational probability distributions
in three dimensions, against the exact orientational probability
distribution in three dimensions. The short time approximation works very well at short and intermediate times. As expected, the exact orientational probability distribution
deviates from the short time approximate orientational probability distribution at very long times. 
In the long time domain, our long time approximate form works very well, as it should.
The predictions that 
stem out of our analysis, which are discussed in this section, can be tested against 
simulations and future experiments.  We finally conclude with some discussions in Section VI.


\section{II. Fokker Planck equation for orientational dynamics of an ABP}

The speed of an ABP is fixed but its direction is a vector diffusing on the 
unit sphere\cite{urna}. 
The Fokker Planck equation describing the 
orientational probability distribution of an 
ABP in $(d+1)$ dimensions\cite{fokkerd} is: 


\begin{align}
\frac{\partial P}{\partial t} =
D_R [\frac{1}{\sqrt{det\hspace{.1cm}g}}\frac{\partial{}}{\partial{\xi^i}}({\sqrt{det\hspace{.1cm}g}}{g^{ij}}\frac{\partial{P}}{\partial{\xi^j}})] \nonumber \\
\label{fokkerd}
\end{align}
where we have considered $d$ arbitrary curvilinear coordinates 
$(\xi^1, \xi^2, \xi^3, ...\xi^d)$ 
on the surface of the unit sphere.
$g^{ij}$ is the inverse metric tensor. 

In three dimensions it takes the following familiar form in polar
coordinates:
\begin{align}
\frac{\partial P(\theta,\phi,t) }{\partial t} = 
D_R[\frac{1}{{\sin{\theta}}}\frac{\partial \sin{\theta}}{\partial {\theta}}\frac{\partial P}{\partial \theta}+\frac{1}{{\sin}^2{\theta}}\frac{\partial^2 P}{\partial{\phi}^2}]  \nonumber \\
\end{align}

This Fokker Planck equation admits an exact solution as a Kernel 
in series form\cite{zheng}:
\begin{eqnarray}
K(\theta_0,\phi_0,0,\theta,\phi,t) = \sum_{l=0}^{\infty}\sum_{m=-l}^{l}{e^{-D_Rl(l+1)t}Y_{l}^{m^{*}}(\theta_0,\phi_0)} \nonumber \\ 
{Y_{l}^{m}(\theta,\phi)}
\label{ylm}
\end{eqnarray}
Considering an initial distribution $P(\theta,0)$ located at the north pole of the sphere, the azimuthal symmetry leads to a simpler form for the series solution\cite{fokkerd}: 
\begin{align}
P_{exact}(\theta, t) = \frac{\sin{\theta}}{2}\sum_{l=0}^\infty{(2l+1)e^{-\frac{l(l+1)t}{2}}}P_l(\cos{\theta})
\label{legendre}
\end{align}
where we have included the measure $\sin{\theta}$ and set $D_R = \frac{1}{2}$.
Although the solution in Eq. (\ref{legendre}) is exact, poor convergence at short times is expected to come in the way of an effective 
implementation in the short time domain.


\section{III. Orientational Probability Distribution of an ABP: A Short Time approximation}

In this section we arrive at an effective short time approximate form for the Orientational 
Probability Distribution of an ABP which can be implemented effectively at short and intermediate times. 

The  solution to the Fokker Planck equation (\ref{fokkerd}) on the surface of a $(d+1)$ dimensional unit sphere is given formally by the 
Wiener integral kernel\cite{gaussian}. This kernel $K(\hat{n_1},0,\hat{n_2},\tau)$ is the probability of the particle to be at $\hat{n_2}$ at time $\tau$ given that it was at $\hat{n_1}$ at time $0$.:
\begin{align}
K(\hat{n_1},0,\hat{n_2},\tau) 
= \int{D[\hat{n}(\tau)] exp[-S[\hat{n}(\tau)]]} 
\label{kernel}
\end{align}
where $S[\hat{n}(\tau)]= \frac{1}{2}\int_0^\tau{\frac{d \hat{n}}{d\tau}.\frac{d \hat{n}}{d\tau}{d\tau} }$.

At short times, this kernel is dominated by the classical path connecting $\hat{n_1}$ to $\hat{n_2}$ and is given by\cite{gaussian,meerson}:

\begin{align}
K(\hat{n_1},0,\hat{n_2},\tau)
\sim exp-S_{cl}[\hat{n_1},\hat{n_2},\tau]
\label{kernelclass}
\end{align}
where $S_{cl}[\hat{n_1},\hat{n_2},\tau]$ is the classical action pertaining to the least action path connecting $\hat{n}_1$ and $\hat{n}_2$ in time $\tau$.

Taking into consideration the quadratic fluctuations around the classical path we get\cite{biopolymer},\footnote{In \cite{meerson} the authors derive a short time 
approximate formula using a ``geometric theory of diffusion''. They, however, do not incorporate the quadratic fluctuation corrections that have been taken into consideration in our analysis.}:
\begin{align}
K(\hat{n_1},0,\hat{n_2},\tau)
\sim \sqrt{det V} exp-S_{cl}[\hat{n_1},\hat{n_2},\tau]
\label{kernelapp}
\end{align}
$det V$ is the Van Vleck determinant given by the determinant of the $2\times 2$ Hessian matrix:
\begin{align}
V_{ij}
=\frac{\partial^{2} S_{cl}[\hat{n_1},\hat{n_2},\tau]}{\partial \hat{n_1}^{i} \partial \hat{n_2}^{j}}
\label{hessian}
\end{align}

Finally, incorporating the normalisation ${N}(\tau)$
we arrive at \cite{kleinert,gelfand}:
\begin{align}
K(\hat{n_1},0,\hat{n_2},\tau)
= {N}(\tau)\sqrt{det V} exp-S_{cl}[\hat{n_1},\hat{n_2},\tau]
\label{kernelappnorm}
\end{align}


Varying the action in Eq. (\ref{kernel}) we find the classical path governed by the equation:
\begin{align}
\frac{\partial^{2} \hat{n}}{\partial \tau^2}= \lambda \hat{n}
\label{geodesic}
\end{align}

with $\lambda$ is a Lagrange multiplier enforcing the 
constraint ${\hat n}.{\hat n}=1$. The solution of Eq. (\ref{geodesic}) is the unique great circle passing 
through $\hat{n}^1$ and $\hat{n}^2$
(we are assuming here that $\hat{n}^1$ and $\hat{n}^2$ are not collinear) . 
Thus the classical action is given by: 
$$S_{cl}= \frac{\theta^2}{2\tau}.$$

$\theta= Cos^{-1}(\hat{n}_1 . \hat{n}_2).$

Our fluctuation determinant is a $({d+1})\times ({d+1})$ matrix $\tilde{V}_{ij}$
\begin{align}
\tilde{V}_{ij}
=\frac{\partial^{2} S_{cl} }{\partial {n_1}^{i} \partial {n_2}^{j}}
\label{hessiantilde}
\end{align}

Consider Cartesian coordinates so that $n_1$ and $n_2$ lie in the $x-z$ plane. In addition there are $(d-1)$ transverse dimensions.

We notice that on reflecting in the $x-z$ plane the $({d-1}) \times {2}$ matrix block $\tilde{V}_{a\beta}$ and the ${2} \times ({d-1})$ matrix block $\tilde{V}_{\alpha b}$  
in the $({d+1})\times ({d+1})$ matrix $\tilde{V}$ change sign. The requirement of invariance tells us that the entries in these two blocks are all zero. Thus we are left with
two blocks with nonzero entries: the $2\times2$ matrix $\tilde{V}_{ab}$ pertaining to the $x-z$ plane and the $({d-1})\times({d-1})$ 
matrix block $\tilde{V}_{\alpha \beta}$.
Taking into consideration the rotational invariance of the $({d-1})\times({d-1})$ matrix block $\tilde{V}_{\alpha \beta}$ in $\tilde{V}$, we find that it is of the form
$\tilde{V}_{\alpha \beta}= \tilde{V} \delta_{\alpha \beta}$.
Thus the determinant pertaining to the $(d+1) \times (d+1)$ matrix $V$ is $det V = {det \tilde{V}_{ab}}  {det \tilde{V}_{\alpha \beta}}.$ 





We first compute the determinant corresponding to the $2\times 2$ matrix $\tilde{V}_{ab}$ \footnote{This is effectively a $1 \times 1$ determinant since 
the matrix has a null eigenvalue.}.
This determinant can be expressed as follows: 
\begin{align}
det \tilde{V}_{ab}
= \hat{n}_{1a} \epsilon^{ac} \hat{n}_{2b} \epsilon^{bd} V_{cd}
\label{detvtwo}
\end{align}

An explicit computation of this $2 \times 2$ determinant gives us:
$$det{\tilde  {V}_{ab}} = \frac{1}{\tau}.$$

Thus the total propagator is given by 
\begin{align}
K(\hat{n_1},0,\hat{n_2},\tau)
= N(\tau)\sqrt{\frac{1}{\tau}}exp[-{\frac{{\theta}^2}{2 \tau}}]
\label{kernelapp}
\end{align}

The normalisation $N(\tau)$ can be fixed by integrating the kernel $K(\hat{n_1},0,\hat{n_2},\tau)$ over initial conditions to 
get $P(\theta, \tau)$ and then by numerically imposing the normalisation condition $\int{P(\theta, \tau) d{\theta}}=1$.

The fact that this expression is correct can be seen by 
specialising to $d=1$ and 
noticing that diffusion on a circle can be well approximated by diffusion on a line 
($x$ axis) for short times:

\begin{align}
K= \frac{1}{\sqrt{4 \pi D t}}exp[-\frac{{x}^2}{4 D t}]  
\label{kernelline}
\end{align}

where $D$ is the diffusion constant.

The remaining $(d-1) \times (d-1)$ matrix $\tilde {V_{\alpha \beta}}$   has a diagonal form and corresponds to a determinant $det \tilde {V_{\alpha \beta}}=  (\frac{\theta}{\tau \sin{\theta}})^{(d-1)}$.
Thus the determinant pertaining to the $(d+1) \times (d+1)$ matrix $\tilde{V}$ is $det \tilde{V} = {det \tilde{V}_{ab}}  {det \tilde{V}_{\alpha \beta}}= \frac{1}{\tau}(\frac{\theta}{\tau \sin{\theta}})^{(d-1)} $. 
(See Appendix for a more explicit and algebraic derivation).


As a special case, we can consider the three dimensional determinant which gives us, on setting $d=2$ (the number of angular dimensions in three dimensions): 
$det V^{(3)} = \frac{1}{\tau^2}(\frac{\theta}{ \sin{\theta}}) $.

Thus we get: 

\begin{align}
K(\hat{n_1},0,\hat{n_2},\tau)
= \frac{{N}(\tau)}{\tau}\sqrt{\frac{\theta}{\sin{\theta}}}exp[-{\frac{{\theta}^2}{2 \tau}}] 
\label{kernelthreedapp}
\end{align}

Multiplying the above expression for $K(\hat{n_1},0,\hat{n_2},\tau)$ by the measure $\sin{\theta}$ we arrive at the following approximate
short time probability distribution: 

\begin{align}
P^{S}_{approx}(\theta,\tau)
= \frac{{N}(\tau)}{\tau}\sqrt{\theta \sin{\theta} }exp[-{\frac{{\theta}^2}{2 \tau}}]
\label{probthreedapp}
\end{align}

Notice that in contrast to $P(\theta,t)$ where time $t$ appears in the numerator of the argument of the exponential, the approximate propagator $P^{S}_{approx}(\theta,\tau)$ 
has much better convergence properties at short times because of the appearance of time in the denominator of
the argument of the exponential.


\section{IV. Orientational Probability Distribution of an ABP: A Long Time approximation}

In this section we discuss an approximate analytic form for the Orientational Probability Distribution of an ABP.
This can be obtained simply by noticing that at long times, the expression for the Orientational Probability Distribution 
given by the series in Eq. (\ref{legendre}) is dominated by the first few terms. Truncating the series and retaining the first three terms in the summation, we get:
\begin{eqnarray}
P^{L}_{approx}(\theta,\tau)
= \frac{1}{2}\sin{\theta}[1+3\cos{\theta}e^{-\tau} \\ \nonumber
+\frac{5}{2}(3\cos^2{\theta} -1)e^{-3\tau}]
\label{longtimeapp}
\end{eqnarray}

Thus Eq. (\ref{longtimeapp}) gives us an analytical form for the long time Orientational Probability Distribution.


\section{V. Orientational Probability Distribution of an ABP: Comparison of the exact solution with approximate analytical forms}

In this section, we restrict to three dimensions and graphically compare the exact series solution for the orientational probability distribution of
an ABP with the approximate analytical forms obtained in the short time regime and the long time regime.
We first compare the exact series solution with the
short time approximate solution obtained by using saddle point methods. We notice that at short and intermediate times
the two probability distributions agree very well (Fig. $1$ and Fig. $2$ ). The two probability distributions deviate from one another at long times, as expected (Fig. $3$ ).
Thus, we conclude that the short time approximate form works well at short and intermediate times.

We then compare the the exact series solution with the
long time approximate solution obtained in Sec. IV and the short time
approximate solution obtained in Sec. III.
We notice that even at relatively short times ($t=0.6$), the short time approximate distribution, the long time approximate distribution and 
the exact distribution agree very well (See Fig. $4$). 
At an intermediate time (t=1.5) the short time approximate Probability Distribution, the long time approximate
Probability Distribution and the exact Probability Distribution merge (See Fig. $5$). At long times,
the long time approximation works very well (See Fig, $6$) whereas the short time approximation deviates
considerably from the exact distribution. At very short times (say $t=0.2$), the long time approximate 
distribution shows oscillations, indicating a breakdown of the long time approximation at very short times,
stemming from truncation errors.

We also display a comparison of the three distributions (the short time, the long time and the exact) using
the Kullback Leibler Divergence. This measure, also known as the relative entropy, is widely used in Information 
Theory to measure the extent of deviation between a trial distribution and a fiducial one.
In our case the fiducial distribution is the exact distribution and the trial one is the approximate form (the 
short time approximate form or the long time approximate form, as the case may be).
\begin{eqnarray}
D_{KL}:=\int_0^{\pi}{d{\theta} P_{approx}(\theta) Log[\frac{P_{approx}(\theta)}{P_{exact}(\theta)}]}
\label{kullback}
\end{eqnarray}
In the table below, the relative entropy of the long time approximate distribution with respect to the 
exact distribution is denoted by $D_{KL} long$ and the relative entropy of the short time approximate distribution with respect to the 
exact distribution is denoted by $D_{KL} short.$ 

\begin{center}
\begin{tabular}{ |c|c|c| }
 \hline
 Time & $D_{KL} long$ & $D_{KL} short$ \\
\hline
 0.5 & 0.0732474 & $3.79124 \times 10^-6$ \\
 1.0 & 0.0000489801  & $3.79124 \times 10^-6$ \\
1.5 & $6.87399 \times 10^-8$ & 0.000551454 \\
2.0 & $1.44112 \times 10^-10$  & 0.0009733 \\
3.0 & 0 & 0.00215108 \\
5.0 & 0 & 0.0084298 \\
10.0 & 0 & 0.0272658 \\
 \hline
\end{tabular}
\end{center}
This table shows that the short time approximate analytical form works well at short and intermediate times and the long time approximate
analytical form works well at long times. 
Thus we conclude that we have excellent analytical forms for the ${\it entire}$ range of time scales.

\begin{figure}[h!]
                \begin{center}
                        \includegraphics[width=0.4\textwidth]{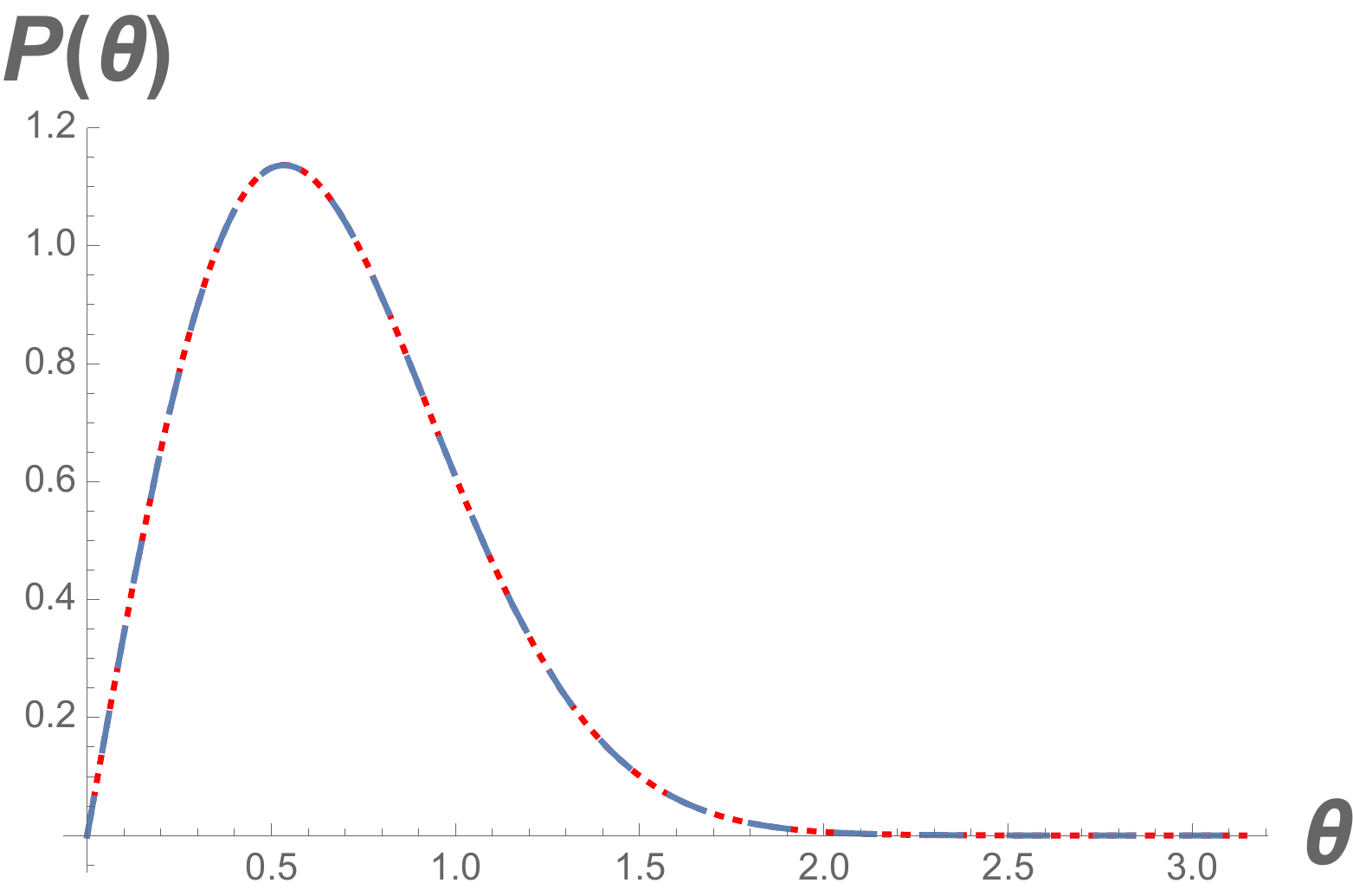}
                        \caption{Comparison of the short time approximate 
Probability Distribution (Eq.(\ref{probthreedapp}))(red dotted line) and the exact Probability Distribution (Eq. (\ref{legendre}))(blue dashed line) at time $t=0.3$.}
                \end{center}
        \end{figure}

\begin{figure}[h!]
                \begin{center}
                        \includegraphics[width=0.4\textwidth]{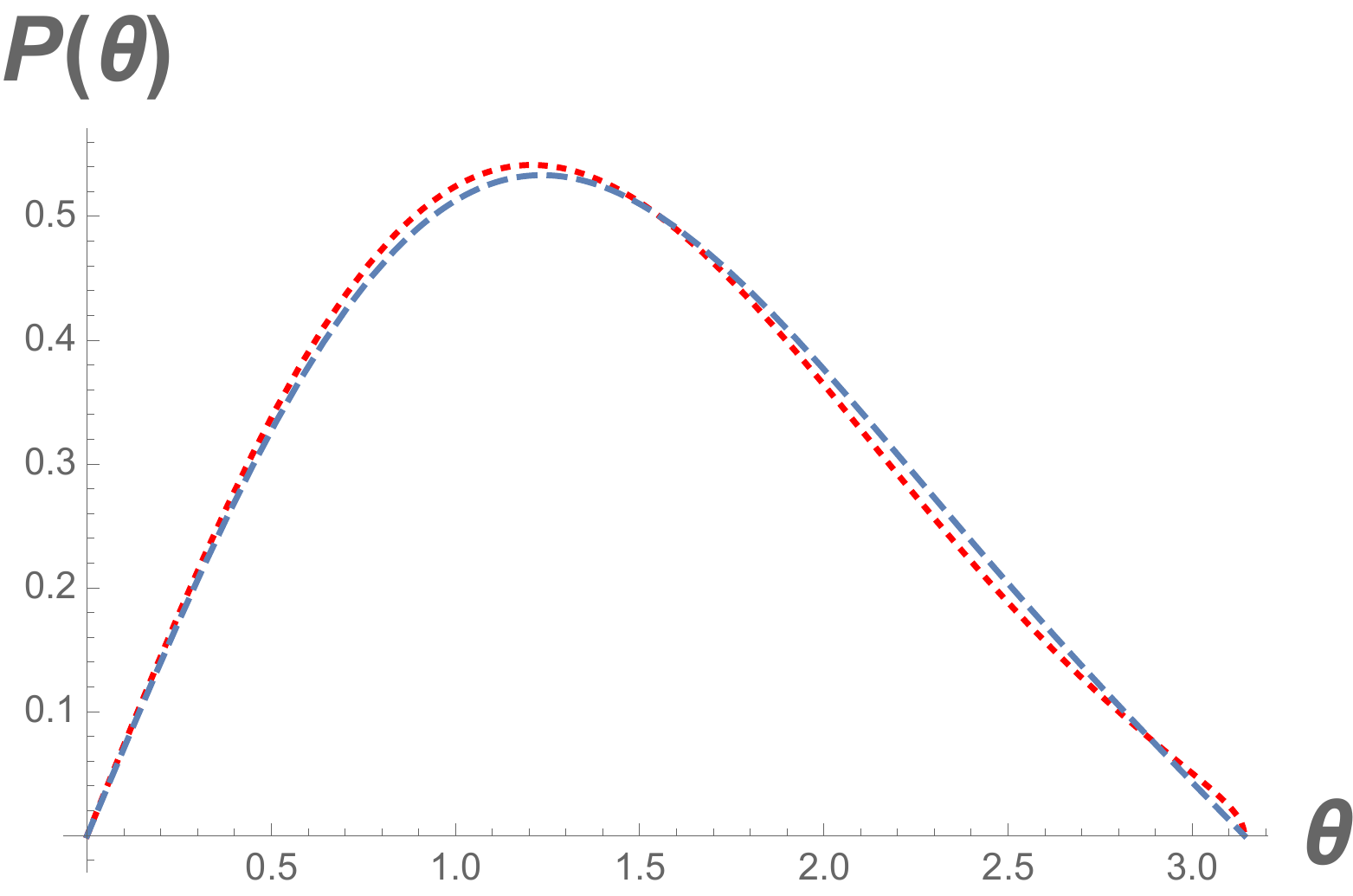}
                        \caption{Comparison of the short time approximate Probability Distribution (Eq.(\ref{probthreedapp}))(red dotted line) and the exact Probability 
Distribution (Eq.(\ref{legendre}))(blue dashed line) at time $t=2.$}
                \end{center}
        \end{figure}

\begin{figure}[h!]
                \begin{center}
                        \includegraphics[width=0.4\textwidth]{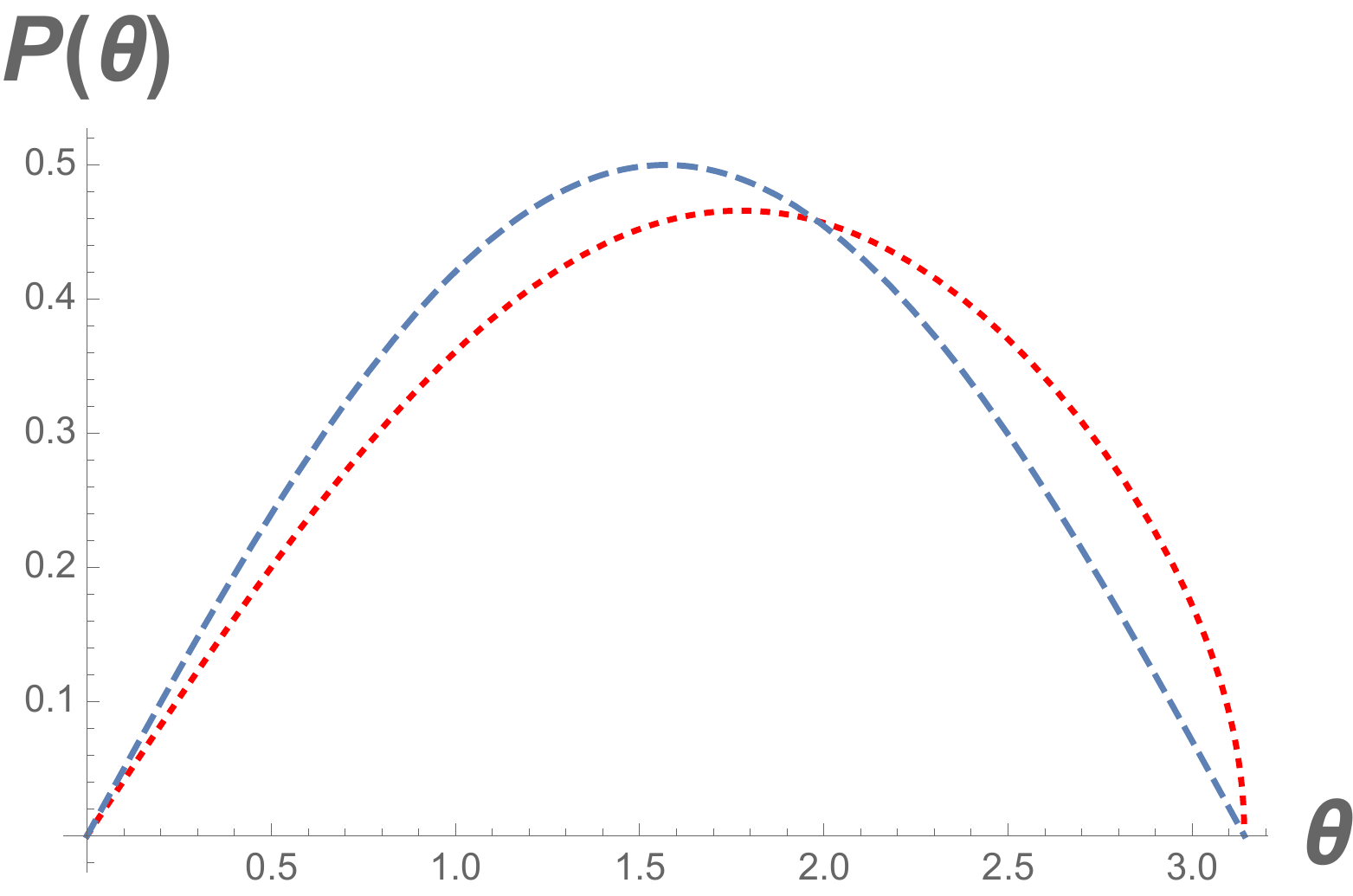}
                        \caption{Comparison of the short time approximate Probability Distribution (Eq.(\ref{probthreedapp}))(red dotted line) 
and the exact Probability Distribution (Eq.(\ref{legendre}))(blue dashed line) at time $t=10$.}                       
                \end{center}
        \end{figure}

\begin{figure}[h!]
                \begin{center}
                        \includegraphics[width=0.4\textwidth]{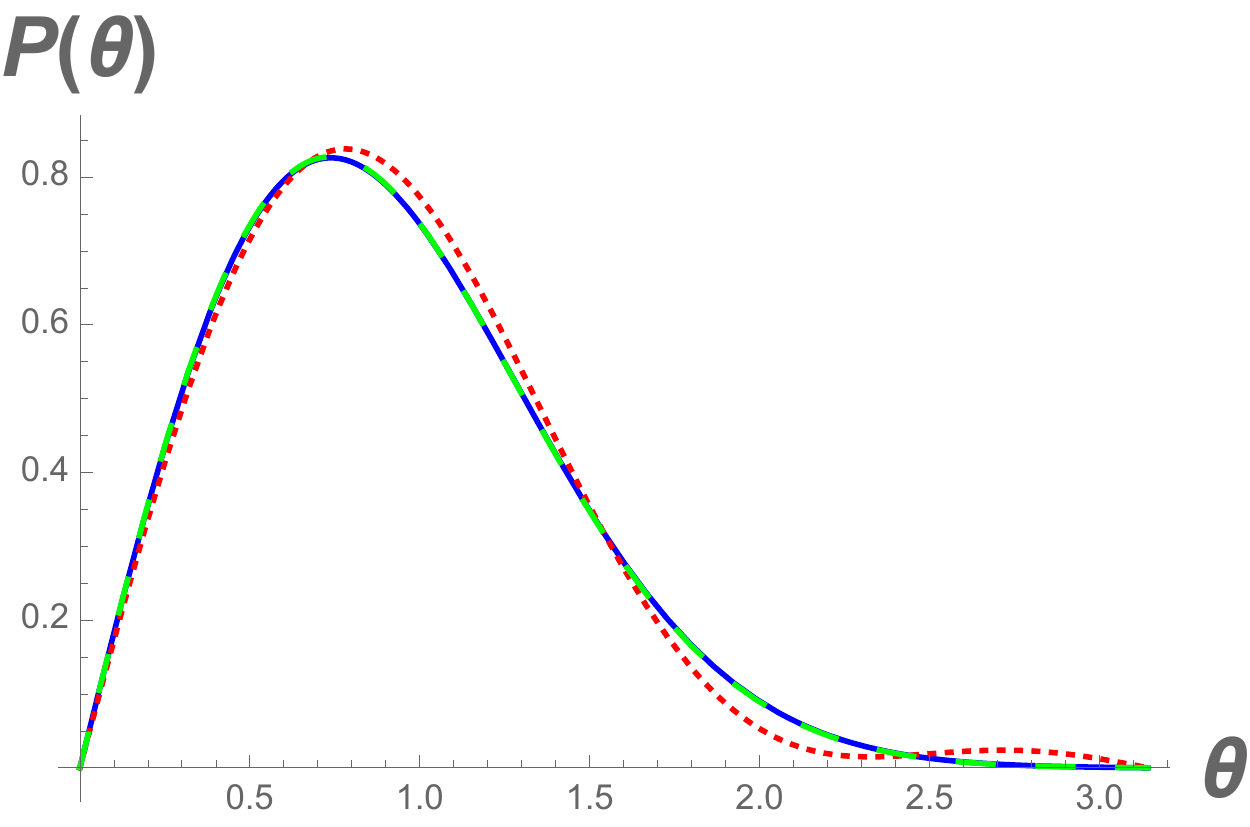}
                        \caption{Comparison of the short time approximate 
Probability Distribution (Eq.(\ref{probthreedapp}))(green dashed line), the exact Probability Distribution (Eq.(\ref{legendre})) (blue line) and 
the long time approximate Probability Distribution (Eq.(\ref{longtimeapp}) (red dotted line) at time t=0.6.}
                \end{center}
        \end{figure}
\begin{figure}[h!]
                \begin{center}
                        \includegraphics[width=0.4\textwidth]{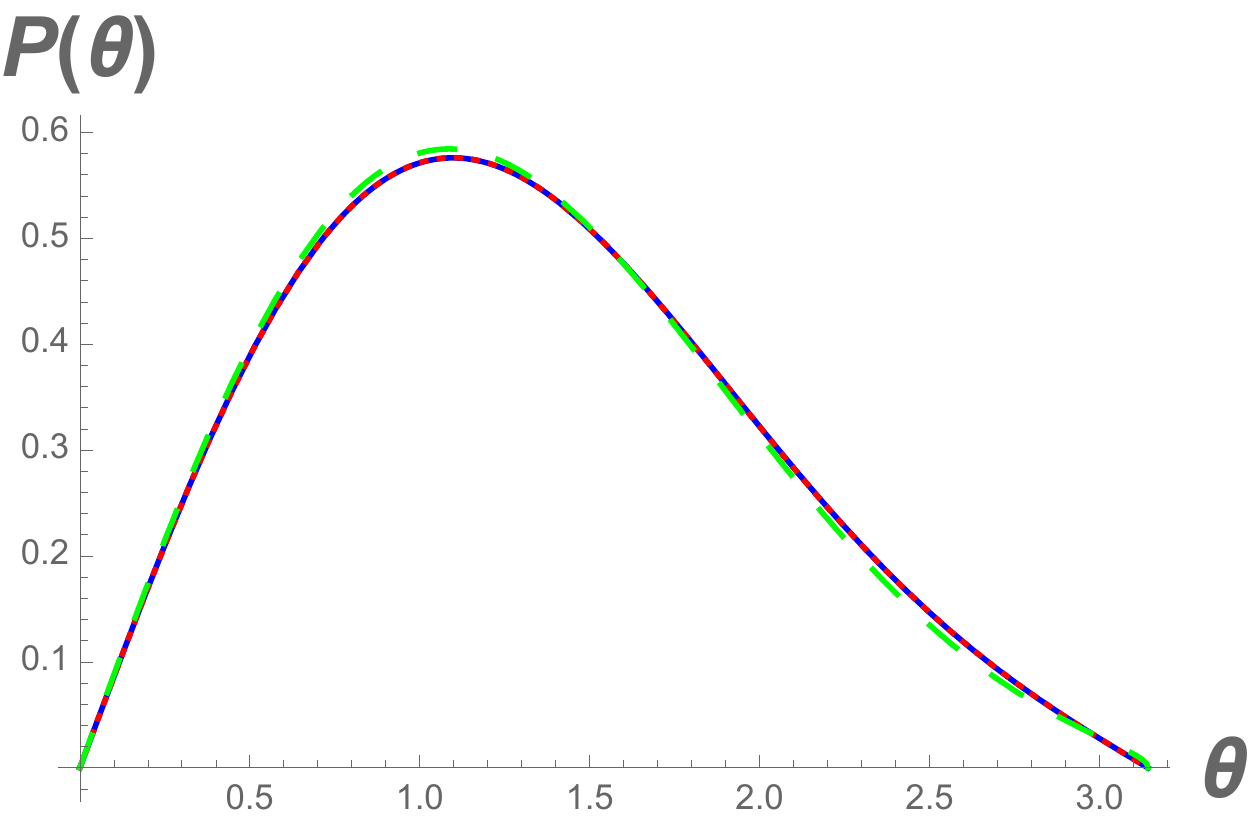}
                        \caption{Comparison of the short time approximate
Probability Distribution (Eq.(\ref{probthreedapp}))(green dashed line), the exact Probability Distribution (Eq.(\ref{legendre})) (blue line)and 
the long time approximate Probability Distribution (Eq.(\ref{longtimeapp}) (red dotted line) at time t=1.5.}
                \end{center}
        \end{figure}
\begin{figure}[h!]
                \begin{center}
                        \includegraphics[width=0.4\textwidth]{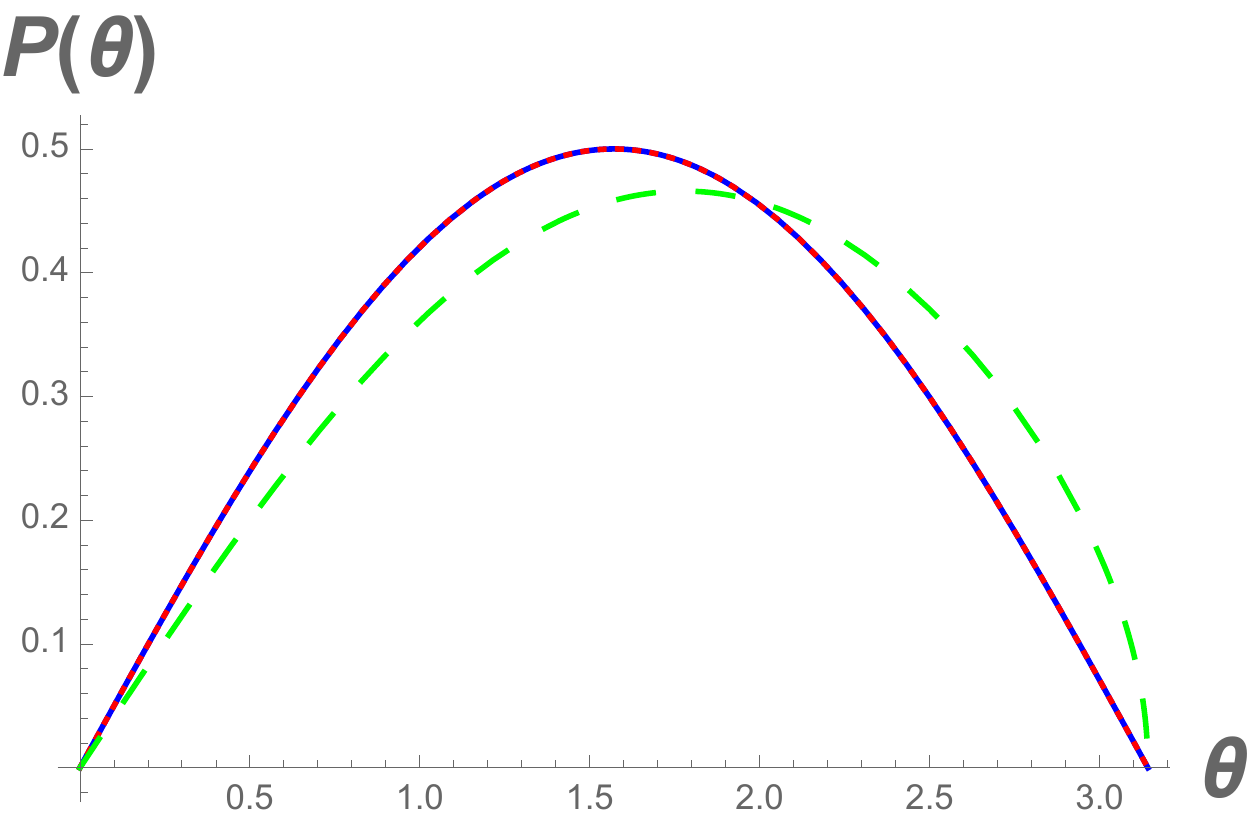}
                        \caption{Comparison of the short time approximate
Probability Distribution (Eq.(\ref{probthreedapp}))(green dashed line), the exact Probability Distribution (Eq.(\ref{legendre})) (blue line)and
the long time approximate Probability Distribution (Eq.(\ref{longtimeapp}) (red dotted line) at time t=10.}
                \end{center}
        \end{figure}

\section{VI.  Conclusion}

We study the orientational probability distribution of an ABP via the 
Fokker Planck equation. Our starting point is the orientational Fokker Planck equation 
of an Active Brownian Particle (ABP) in $(d+1)$ dimensions.
It is well known that the Fokker Planck equation for orientational
dynamics admits an exact solution in series form
which, however, has poor convergence for short and intermediate times.

We present an analytical closed form expression,
which gives a good short time approximate orientational probability distribution.
In some earlier papers\cite{abhisemi,abhinew}
methods used in the study of equilibrium properties of semiflexible polymers have been
incorporated to study Active Brownian Particle dynamics. Here we use such methods to analyse the orientational  probability distribution of an
Active Brownian Particle at short and intermediate times.
The analytical formula is derived using saddle point methods for short times.
In computing the approximate orientational probability distribution we have
considered a situation where the particle winds around the great circle once. We have neglected higher order
windings. The excellent agreement between our approximate probability distribution and the exact
probability distribution shows that higher order windings are negligible and our restriction to
only one winding is indeed a very good approximation.
We also present a long time approximate analytical form for the 
orientational probability distribution of an ABP. Thus we have presented explicit analytical forms for the orientational probability distribution
of an ABP in the ${\it entire}$ range of time scales.  

Restricting to three dimensions, we graphically compare the exact formula with the short time and the long time approximate probability distributions and show 
that the short time approximate formula works 
well at short as well as at intermediate times. The short time approximate formula deviates from the exact one at long times, as expected.
In the long time domain, our long time approximate analytical expression for the orientational probability distribution works well. 
Our predictions can be tested against future experiments and simulations probing orientational probability distribution of an ABP.

\section{Acknowledgements}
We acknowledge Urna Basu, Abhishek Dhar  and Satya Majumdar for discussions. We thank Baruch Meerson for drawing our attention to Ref\cite{meerson}, which deals with the
``geometric theory of diffusion''.

\section{Appendix : Computation of the Van Vleck determinant}
Here we present an explicit computation of the Van Vleck determinant, which has been schematically outlined in the main body of the paper. 
We use explicit notation, writing $\hat{r_1}$ and $\hat{r_2}$ for  $\hat{n_1}$ and $\hat{n_2}$
and setting $r$ to $n$ after differentiation. Consider two vectors $\hat{r}^L(\gamma)$ and $\hat{r}^T(\psi)$. $\hat{r}^L$ lies in the $\hat{x}-\hat{z}$ plane and $\hat{r}^T$ is 
in the transverse direction. 
We consider variations of $\hat{n}$ in the transverse direction described by
$\hat{r}^T$
\begin{eqnarray}
\hat{r}^T(\psi)
= (\cos{\psi}) \hat{n} + (\sin{\psi}) \hat{y}^\alpha
\label{rlongtrans}
\end{eqnarray}
(where $\alpha=1,2,3,.....,(d-1)$)
and variations in the $x-z$ plane described by $\hat{r}^L$.
\begin{eqnarray}
\hat{r}^L(\gamma)
= (\cos{\gamma}) \hat{x} + (\sin{\gamma}) \hat{z}
\label{rlongtrans}
\end{eqnarray}

We first compute the determinant of the $2 \times 2$ matrix in the $x-z$ plane.

Consider $\hat{r}_1^{L}$ and $\hat{r}_2^{L}$ which can be expressed as follows in the $x-z$ plane:
\begin{eqnarray}
\hat{r}_1^{L}(\gamma_1)
= \cos{\gamma_1} \hat{x} + \sin{\gamma_1} \hat{z} 
\label{r1vec}
\end{eqnarray}
\begin{eqnarray}
\hat{r}_2^{L}(\gamma_2)
= \cos{\gamma_2} \hat{x} + \sin{\gamma_2} \hat{z}                                 
\label{r2vec}
\end{eqnarray}
The dot product $\hat{r}_1^{L}.\hat{r}_2^{L}= \cos(\theta)$
where $\theta= {(\gamma_1 - \gamma_2})$.

Thus the $1\times 1$ determinant 
$\frac{\partial^{2} S_{cl} }{\partial {\gamma_1} \partial {\gamma_2}}|_{\gamma_1=\theta_1, \gamma_2=\theta_2}= -\frac{1}{\tau} $
	
	 
Let us now consider 
$\tilde{V}_{a\beta}$ and 
$\tilde{V}_{\alpha b}$, the off diagonal matrix blocks of the 
$({d+1})\times ({d+1})$ matrix $\tilde{V}$.

The Van Vleck determinant pertaining to $\tilde{V}_{a\beta}$ can be computed as follows.
Consider 
\begin{eqnarray}
\hat{r_1}^T(\psi_1)
= (\cos{\psi_1}) \hat{n_1} + (\sin{\psi_1}) \hat{y}^{\alpha}
\label{rlongtrans}
\end{eqnarray}
where $\alpha =1,2,3,.....,(d-1)$.
\begin{eqnarray}
\hat{r_2}^L(\gamma_2)
=(\cos{\gamma_2}) \hat{x} + (\sin{\gamma_2}) \hat{z}
\label{r1vec}
\end{eqnarray}


Notice that$\frac{\partial S_{cl} }{\partial {\psi_1}}= -\frac{\theta}{\tau}
\frac{\partial \theta}{\partial {\psi_1}} $.
Taking the dot product of $\hat{r_1}^T(\psi_1)$ and $\hat{r_2}^L(\gamma_2)$ and 
taking the derivative with respect to $\psi_1$ we get:
\begin{eqnarray}
-\sin{\theta}\frac{\partial{\theta}}{\partial{\psi_1}}
= (-\sin{\psi_1})(\cos{\gamma_2})({\hat{n}_1}.{\hat{x}})\\ \nonumber
-(\sin{\psi_1})(\sin{\gamma_2}) ({\hat{n}_1}.{\hat{z}})    
\label{dotproductder}
\end{eqnarray}

Clearly, computation of the Van Vleck determinant gives us
\begin{eqnarray}
\frac{\partial^2 {S_{cl} }}{ \partial{\gamma_2} \partial{\psi_1} }|_{\psi_1 =0,\gamma_2=\theta_2}
= 0
\label{vanvleck}
\end{eqnarray}
in accord with the symmetry argument presented in the main body of the paper. 
Following similar steps it can be shown that the Van Vleck determinant 
pertaining to $\tilde{V}_{\alpha b}$ is: 
\begin{eqnarray}
\frac{\partial^2 {S_{cl} }}{ \partial{\gamma_1} \partial{\psi_2} }|_{\psi_2 =0, \gamma_1=\theta_1}
= 0
\label{vanvleckother}
\end{eqnarray}	
	
Finally we compute the determinant of the remaining $({d-1})\times ({d-1})$ matrix block $\tilde{V}_{\alpha \beta}$ in $\tilde{V}$.
	
Consider two vectors $\hat{r_1}^T$ and $\hat{r_2}^T$. 

\begin{eqnarray}
\hat{r_1}^T(\psi_1)
= (\cos{\psi_1}) \hat{n_1} + (\sin{\psi_1}) \hat{y}^{\alpha}
\label{rtransone}
\end{eqnarray}
\begin{eqnarray}
\hat{r_2}^T(\psi_2)
= (\cos{\psi_2}) \hat{n_2} + (\sin{\psi_2}) \hat{y}^{\beta}
\label{rtranstrans}
\end{eqnarray}

As before, taking the dot product of $\hat{r_1}^T(\psi_1)$ and $\hat{r_2}^T(\psi_2)$ and taking the derivative with respect to $\psi_1$ we get:	
\begin{eqnarray}
-\sin{\theta}\frac{\partial{\theta}}{\partial{\psi_1}}
= (-\sin{\psi_1})(\cos{\psi_2})({\hat{n}_1}.{\hat{n}_2})\\ \nonumber
+(\cos{\psi_1})(\sin{\psi_2}) ({\hat{y}^{\alpha}}.{\hat{y}^{\beta}})
\label{dotproductder}
\end{eqnarray}

Now 
\begin{eqnarray}
\frac{\partial {S_{cl} }}{ \partial{\psi_1} }
= \frac{\theta}{\tau}  \frac{\partial \theta}{ {\partial \psi_1} } 
=-\frac{\theta}{\tau}\frac{1}{\sin{\theta}}[(-\sin{\psi_1}\cos{\psi_2})({\hat{n}_1}.{\hat{n}_2})\\ \nonumber
+(\cos{\psi_1})(\sin{\psi_2}) ({\hat{y}^{\alpha}}.{\hat{y}^{\beta}})] 
\label{firstder}
\end{eqnarray}

Thus we finally get:
\begin{eqnarray}
\tilde{V}_{\alpha \beta}
= \frac{\theta}{\tau}\frac{1}{\sin{\theta}}\delta^{\alpha \beta}
\label{vanvlecktranstrans}
\end{eqnarray}
	
Each diagonal entry of the $({d-1})\times ({d-1})$ matrix block $\tilde{V}_{\alpha \beta}$ is 
$\frac{\theta}{\tau}\frac{1}{\sin{\theta}}$
which gives the final expression for the determinant of $\tilde{V}_{\alpha \beta}$ as
$({\frac{\theta}{\tau}\frac{1}{\sin{\theta}}})^{(d-1)}$.

\bibliography{active}

\end{document}